\documentclass[acmsmall,screen]{acmart}
\AtBeginDocument{%
  }
\usepackage{amsmath}
\usepackage{graphicx}
\usepackage{algorithm}
\usepackage{algpseudocode}
\usepackage{textcomp}
\usepackage{xcolor}
\usepackage{multirow}
\usepackage{framed}
\usepackage[framemethod=TikZ]{mdframed}
\usepackage{mdframed}
\usepackage{booktabs}
\acmConference[Conference acronym 'XX]{Make sure to enter the correct
  conference title from your rights confirmation email}{June 03--05,
  2018}{Woodstock, NY}

\begin{document}
\title{CogAdapt: Cognition-informed Sparse Adaptation of Code LLMs}

\author{Yueke Zhang}
\affiliation{%
  \institution{Vanderbilt University}
  \city{Nashville}
  \state{Tennessee}
  \country{USA}
}

\author{Zihan Fang}
\affiliation{%
  \institution{Vanderbilt University}
  \city{Nashville}
  \state{Tennessee}
  \country{USA}
}

\author{Kevin Leach}
\affiliation{%
  \institution{Vanderbilt University}
  \city{Nashville}
  \state{Tennessee}
  \country{USA}
}

\author{Yu Huang}
\affiliation{%
  \institution{Vanderbilt University}
  \city{Nashville}
  \state{Tennessee}
  \country{USA}
}

\renewcommand{\shortauthors}{Zhang et al.}

\begin{abstract}

Large language models (LLMs) have become increasingly capable of generating code.
However, achieving stronger code-generation performance still often relies on costly model adaptation, i.e., fine-tuning pretrained model parameters.
Meanwhile, prior studies have shown meaningful correspondence between how humans process code and how neural models allocate attention or internal computation.
Building on this relationship, recent human-aligned learning approaches have used cognitive signals, such as attention or brain activity, to guide training and improve model performance.
However, these methods typically still adapt a large portion of the model, leaving the training cost largely unchanged.
This reveals an important gap: human cognitive signals may not only indicate \emph{what} the model must learn from, but also help determine \emph{where} adaptation is most useful within the model.
Motivated by this insight, we investigate whether human responses during code reading correspond to the internal behavior of code models, and whether this correspondence can guide more selective and efficient model adaptation without sacrificing performance.

We present \textbf{CogAdapt}, a cognition-informed framework for task-dependent sparse adaptation of code models.
CogAdapt first learns transferable program-level and token-level priors from human Electroencephalography (EEG) and attention data, then combines these priors with the frozen model's response to each coding task to determine how much adaptation to allocate and which transformer blocks should receive updates.
During fine-tuning, only the selected blocks are updated, while no new human recordings are required for inference.
Across Qwen and GLM, we find consistent correspondence between human reading behavior and MoE computation.
CogAdapt achieves the best pass@1 across both LiveCodeBench and BigCodeBench, including gains of \textbf{10.86} and \textbf{6.29} percentage points over matched regular fine-tuning on LiveCodeBench, while reducing gradient-eligible adaptation parameters by \textbf{86.21--87.21\%}.
These results suggest that human comprehension signals can provide useful guidance for making code-model adaptation both more selective and more effective.

\end{abstract}


\maketitle

\section{Introduction}

Large Language Models (LLMs) have become increasingly capable of generating code and supporting a wide range of programming tasks~\cite{codex,codegen}.
However, these advances come with substantial computational cost.
Training and deploying large generative models require considerable GPU resources and electricity, and recent studies have raised growing concerns about their energy consumption~\cite{strubell,devries2023energy,luccioni2024power}.
Mixture-of-Experts (MoE) architectures provide a promising way to improve computational efficiency by activating only a subset of model components for each input, allowing model capacity to grow while reducing the computation required during inference~\cite{sparsemoe,switch,mixtral,deepseekv2}.
Nevertheless, sparse inference does not eliminate the cost of training and adapting these large models.
MoE models still require substantial computation during training~\cite{deepseekv3,pan2024dense}.
Thus, an important challenge remains: while MoE models can selectively allocate computation during inference, how can we similarly make \emph{learning} more selective during adaptation?

Existing parameter-efficient methods address part of this problem by reducing how many parameters are updated.
Parameter-efficient techniques such as LoRA reduce the number of trainable parameters or the memory required for adaptation~\cite{lora,qlora,adalora}.
However, these methods determine adaptation largely from properties of the model itself, providing limited external guidance about \emph{where} learning would be most useful.
In contrast, the human brain operates under strong metabolic constraints, which have been associated with energy-efficient neural coding and computation~\cite{attwell2001energy,levy2021communication}.
This contrast motivates a different question: can human cognitive signals provide prior knowledge about where computation is most valuable, and use that knowledge to guide which parts of an LLM should be adapted?

\begin{figure*}[t]
    \centering
    \includegraphics[width=0.8\textwidth]
    {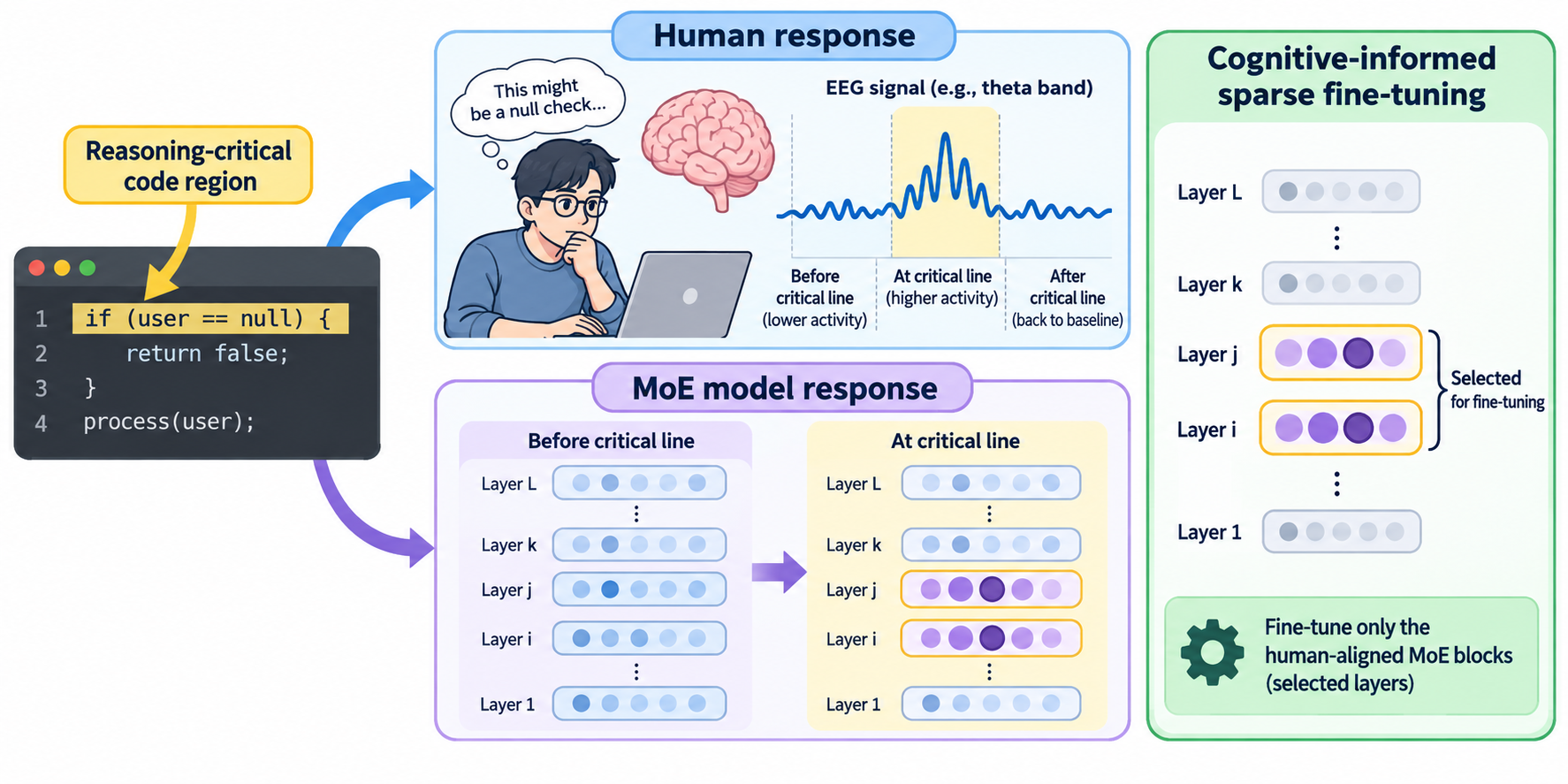}
    \caption{Human attentional hotspots coincide with MoE computation at the same code regions
    }
    \label{fig:hook}
\end{figure*}

Human cognition therefore provides a promising source of prior knowledge for more selective model adaptation.
Eye-tracking studies show that developers follow selective and non-linear reading patterns and allocate different amounts of attention to different code regions~\cite{bansal2023humanattention,wallace2025programmer}.
A growing line of work has consequently modeled programmer attention from eye movements, aligned human reading behavior with neural code representations, and incorporated human attention into code models~\cite{bansal2023humanattention,zhang2024eyetrans, alakmeh2024predicting,wallace2025programmer,zhang2026stop}.
More recent studies extend this direction to Code LLMs by transferring cognitive signals into downstream model adaptation~\cite{zhang2025programmerattention,zhang2025codeactr, zhang2026scancoder,zhang2026eyemulator,zhang2026eyelayer}.
Together, these studies suggest that human cognitive signals capture information about where developers concentrate their processing effort, and that this information can provide useful supervision beyond conventional input--output training data.
However, balancing the performance benefits of human-guided adaptation with training efficiency remains an open challenge.
This motivates our work: can human EEG and attention measurements collected during code comprehension provide useful guidance about which parts of a code
model should be updated? If so, fine-tuning could focus on a small, task-dependent subset of transformer blocks rather than adapting the model uniformly.

We propose \emph{cognition-informed sparse adaptation} of code LLMs, called \emph{CogAdapt}.
Our approach uses EEG and attention measurements from human code reading to inform both the emphasis of code supervision and the allocation of adaptation gradients.
Figure~\ref{fig:hook} illustrates the local human--model correspondence that motivates the study.
We structure our study in two connected steps.
First, we conduct an exploratory analysis to examine whether human attention and EEG responses systematically correspond to MoE computation at the same code regions and across model depth.
The resulting correspondence is distributed across blocks rather than concentrated in one fixed location, motivating task-dependent adaptation.
Building on this finding, CogAdapt converts human signals into transferable priors and combines them with frozen-model responses to dynamically select a small set of transformer blocks for each training task.
Only these selected blocks are adapted, while inference requires no EEG or attention input.

Our contributions are as follows:

\begin{itemize}
    \item \textbf{We present \textsc{CogAdapt}, a cognition-informed framework for sparse adaptation of code models.}
    To the best of our knowledge, this is the first work to use human code-comprehension signals to guide \emph{where} a code model should adapt. \textsc{CogAdapt} transfers cognition-derived priors to downstream tasks \textbf{without requiring new human data}.

\item \textbf{We identify local correspondence between human reading behavior and MoE computation.}
Across two MoE architectures and four code categories,
\textbf{75.0--83.3\% of analyzed regions show positive correlations}
between human attention or EEG responses and model computation.

\item \textbf{We show that human processing demand corresponds to MoE computation across model depth.}
The depth--time heatmaps reveal systematic but architecture-dependent relationships between EEG responses and model computation, with different internal signals dominating in Qwen and GLM.

    \item \textbf{We demonstrate that cognition-informed sparse adaptation improves code generation with fewer gradient-eligible parameters.}
    On LiveCodeBench, \textsc{CogAdapt} improves pass@1 over matched all-block fine-tuning by \textbf{10.86} points for Qwen and \textbf{6.29} points for GLM, while reducing adaptation parameters by \textbf{86.21--87.21\%}.
\end{itemize}

\section{Motivating Example}

\begin{figure*}[t]
    \centering
    \includegraphics[width=0.9\textwidth]
    {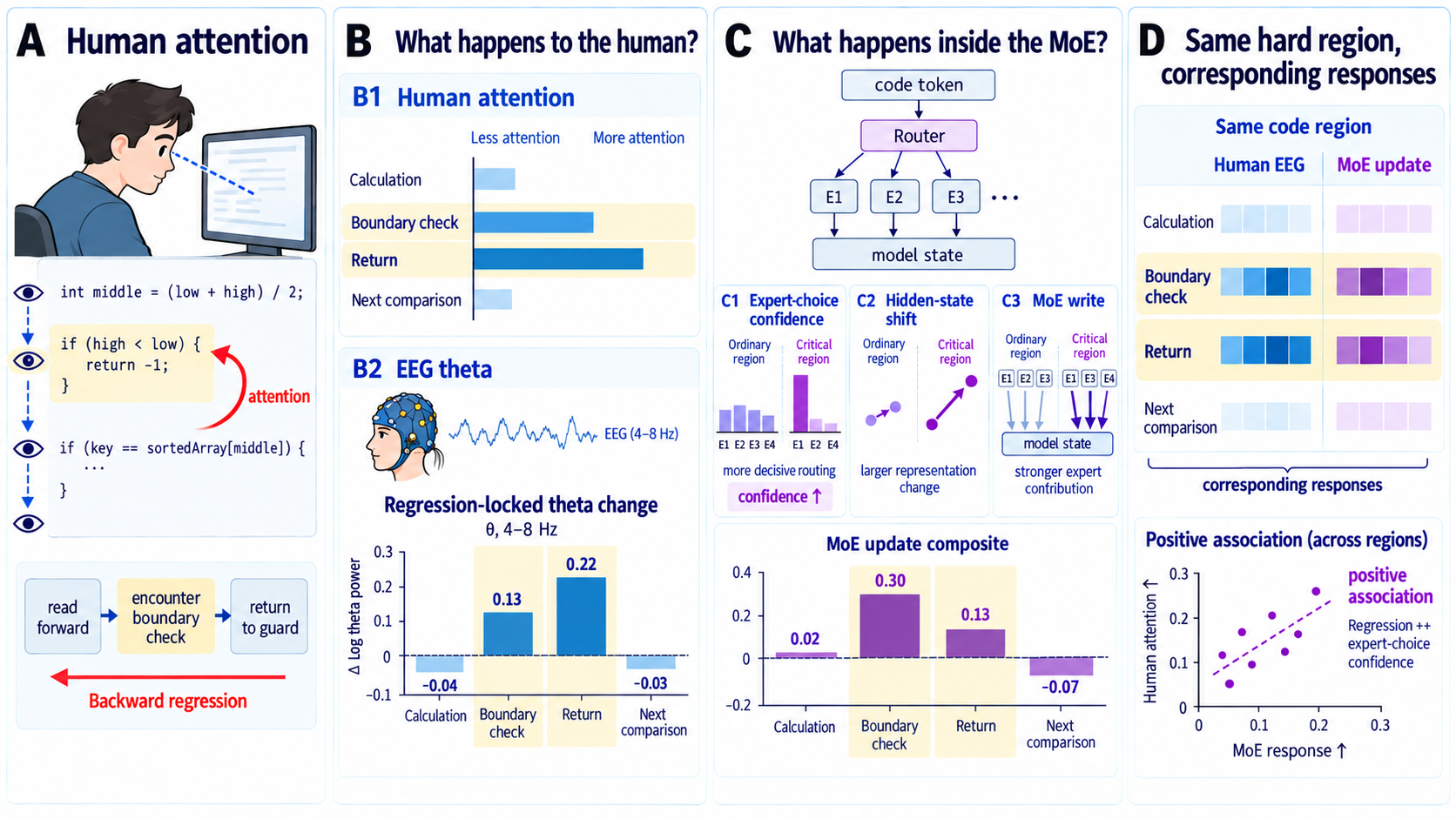}
\caption{A motivating example of human--model correspondence during code comprehension.
(A) A developer looks back at the boundary check while reading the binary-search program.
(B) The same region receives greater human attention and a stronger EEG response.
(C) When the MoE model processes this code, it also responds more strongly around the boundary check: the model makes a more decisive expert choice, changes its internal representation more, and receives a larger contribution from its expert components.
(D) The same reasoning-intensive code region motivates our broader human--model correlation analysis.
}
    \label{fig:motivation}
\end{figure*}

Figure~\ref{fig:motivation} illustrates the intuition behind our study using a real code example from the NoviceVsExpert dataset~\cite{efficacy}, where participants read Java programs while their EEG and eye movements were recorded.
The example is a short binary-search implementation, but different parts of the program require different levels of reasoning.
In particular, the boundary check \texttt{if (high < low)} determines when the search should stop, and the following \texttt{return -1} handles this case.
Understanding these lines requires connecting the current search range with the control flow that follows.

\paragraph{Humans devote more attention to difficult code regions.}
Panel~A illustrates the developer's eye movements while reading the code.
After initially directing their gaze to later lines in the program, the
developer returns to the boundary check before continuing. This backward
eye movement indicates that the boundary condition receives additional
attention rather than being inspected only once during the initial reading.

Panel~B shows the corresponding EEG responses. The boundary check and return
statement receive more visual attention than the surrounding calculation and
comparison. The EEG measurements show a similar pattern: brain activity is
stronger around these same reasoning-intensive lines. Together, these signals
suggest that the boundary condition is a meaningful point of increased human
processing demand.

\paragraph{The model also responds more strongly around the same region.}
Panel~C shows what happens when an MoE code model processes the same code.
Different tokens can cause different levels of internal computation. Around the
boundary check, the model makes a more decisive routing choice, changes its
internal representation more strongly, and receives a larger contribution from
its expert components. We summarize these responses in the lower part of the
panel, where the boundary check and return again stand out from the surrounding
statements.

\paragraph{Human and model responses highlight similar code regions.}
Panel~D places the two views side by side. The goal is not to claim that human
brain activity and model computation are identical. Rather, the example shows
that the same code region elicits relatively stronger responses in both the
developer and the model. Developers devote more
attention and neural processing to the boundary condition, while the model also
exhibits a stronger internal response there. Across code regions, this produces
a positive association between human attention and model response.

This example motivates the central question of our work: \emph{if difficult
code regions consistently attract both stronger human processing and stronger
model computation, can human comprehension signals help guide where and how a
code model should adapt?} CogAdapt investigates this question across programs
and model architectures, and then uses the resulting human-derived signals to
guide sparse adaptation for downstream code generation.

\section{Related Work}
\label{sec:related_work}

Our work connects three lines of research: learning to generate code,
measuring human program comprehension, and comparing human and model
processing. 

\subsection{Large Language Models for Code}
\label{sec:rw_llm_code}
Early pretrained models developed
representations that connect code with its descriptions~\cite{codebert,codet5,codet5plus}. 
Generative code models extend this foundation by learning to produce
programs from specifications and surrounding context. Research on Codex
and CodeGen established the potential of large-scale code training for
function synthesis and multi-turn programming~\cite{codex,codegen}.
Infilling models and code-specialized foundation models further support
editing within existing programs and conditioning on longer code
contexts~\cite{incoder,codellama}.
Efficient adaptation raises a further question: where should a limited
training budget be spent? Low-rank adaptation restricts learning to
compact weight updates, and adaptive rank allocation distributes that
budget according to parameter importance~\cite{lora,adalora}.
Mixture-of-Experts (MoE) code models offer a related form of selectivity
by activating only some experts for each token~\cite{deepseekv2}.
Selective adaptation has also been studied within parameter-efficient
and layer-wise training. LISA samples layers for optimization,
LoRA-drop uses adapter outputs to assess importance, and CKA-based
layer selection uses representational change to identify adaptation
locations~\cite{pan2024lisa,zhou2025lora,ogawa2026layerwise}.

\subsection{Measuring Human Code Comprehension}
\label{sec:rw_neuroscience_code}


Neuroimaging research treats program comprehension as a cognitive
activity that can be studied beyond completion time and correctness.
Early functional magnetic resonance imaging (fMRI) studies demonstrated
the feasibility of observing brain activity during source-code
comprehension, and subsequent studies examined its relation to working
memory, attention, familiarity, and code review~\cite{icse2014,heads,floyd,ivanova,liu}.
Studies of neural efficiency and cortical representations examine how
familiarity and programming skill relate to the processing of source
code~\cite{efficiency,ikutani}. Electroencephalography (EEG) provides
complementary evidence through changes in brain activity associated with
programming experience and mental workload~\cite{crk,kosti}.

Human measurements also help assess what makes particular code difficult
to understand. Neuroimaging studies of code-complexity metrics examine
whether structural properties capture comprehension demands, while work
on identifier naming and readability measures the cognitive consequences
of how code is presented~\cite{complexity,fakhoury}. Physiological
measurements have additionally been used to estimate difficulty during
software-development tasks~\cite{fritz}. 
Combining modalities helps connect overall effort with the code that
readers inspect. Simultaneous fMRI and eye tracking, gaze-linked
near-infrared measurements, and combined EEG and eye tracking provide
complementary views of the same comprehension
activity~\cite{simultaneous,fakhoury,efficacy}.

\subsection{Cognitive Guidance for Language and Code Models}
\label{sec:rw_neuroscience_models}

Neuroscience provides empirical benchmarks for characterizing the
representations learned by language models. Neural predictivity is
associated with next-word prediction performance, while brain--model
correspondence varies across training stages and representational
depths~\cite{schrimpf,caucheteux2022}. These findings motivate examining
multiple model layers and processing signals. 
Human attention also offers a complementary perspective on language
processing. Comparisons between LLM self-attention and eye-saccade
patterns indicate that scaling improves correspondence with human
reading attention, whereas instruction tuning primarily increases
sensitivity to instructions~\cite{gao2023roles}. 

For code, associations between fMRI recordings, program properties,
and learned representations demonstrate shared information between
human and artificial program processing~\cite{srikant}. Beyond such
comparisons, human attention has become a training signal for code
models. Cognitive simulation based on ACT-R generates code-reading
scanpaths from limited human observations, reducing dependence on
large eye-tracking datasets~\cite{zhang2025codeactr}. Observed visual
attention can also supply token-level loss weights for code-model
fine-tuning~\cite{zhang2026eyemulator}. Integrating simulated scanpaths
with training weights based on cognitive salience and attention order
further supports transfer from C++ reading patterns to Java
tasks~\cite{zhang2026scancoder}. Together, these approaches demonstrate
how observed and simulated reading behavior can inform model
adaptation.

\section{Methodology}
\label{sec:method}
\begin{figure*}[t]
    \centering
    \includegraphics[width=\textwidth]
    {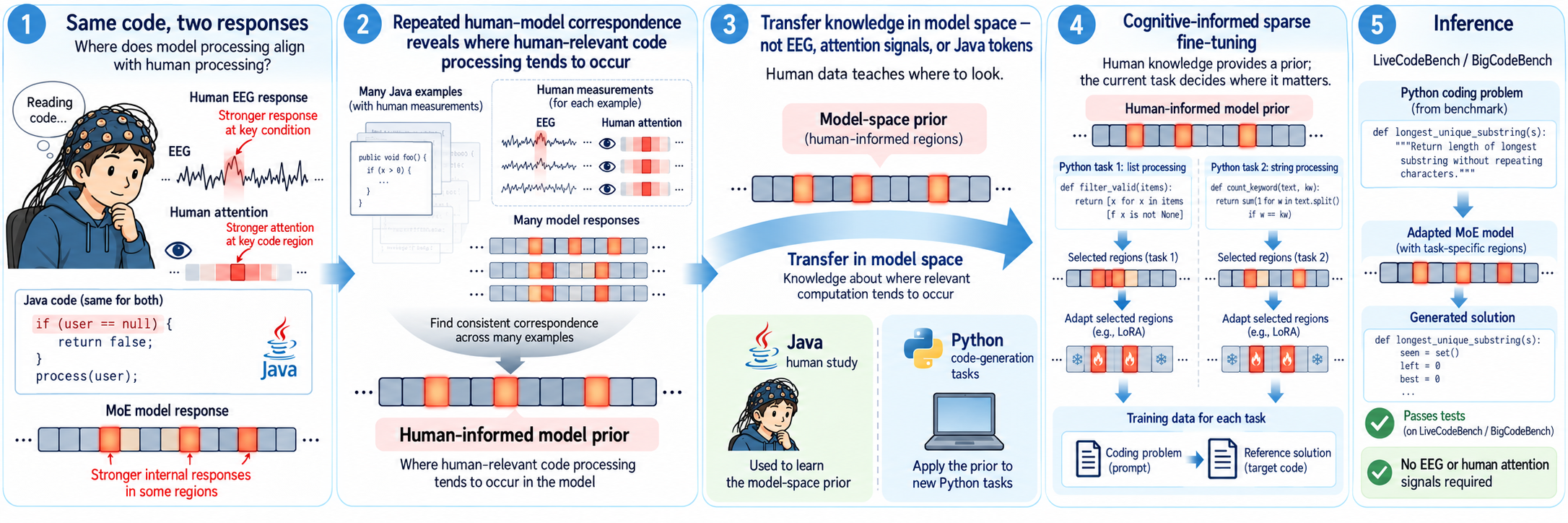}
\caption{Overview of CogAdapt.
We first compare human EEG and attention responses with the internal responses of an MoE code model on the same code.
These human signals are summarized into reusable guidance about which programs and code regions tend to require more processing.
For each downstream coding task, the model then identifies the transformer blocks most relevant for adaptation.
CogAdapt uses this information to focus training on important examples, code regions, and model blocks.
After fine-tuning, inference is unchanged and requires only the coding prompt, with no human measurements.}
    \label{fig:method}
\end{figure*}

We present \textbf{CogAdapt}, a cognition-informed framework for task-dependent sparse adaptation of Mixture-of-Experts (MoE) code models, as shown in Figure~\ref{fig:method}. The central idea is straightforward: human EEG and attention measurements indicate \emph{which code regions require greater cognitive effort during comprehension}, while our human--model analysis reveals \emph{where corresponding changes in model
computation appear across transformer blocks}. CogAdapt transfers this information to downstream code-generation tasks and uses it to determine which training examples and code regions receive greater emphasis and
which transformer blocks are updated.


\subsection{Stage 1: Measuring Human and Model Responses to the Same Code}
\label{sec:method_shared_code}

CogAdapt begins with code for which we have both human-comprehension measurements and model responses.
We use the NoviceVsExpert corpus~\cite{efficacy}, in which participants read Java programs while EEG and eye movements were recorded.
We then process the same programs with each MoE code model.
This gives us two views of the same code: how strongly humans respond while reading it and how strongly different parts of the model respond while processing it.

\paragraph{Human responses.} We obtain two complementary signals from synchronized eye-tracking and EEG recordings of 37 participants reading 32 Java programs~\cite{efficacy}.
Eye tracking identifies which code tokens participants inspect, how long they look at them, and when they look back or revisit them.
From the EEG recordings, we extract theta-band power ($4$--$8$ Hz) and standardize it within participants.
We aggregate theta over the corresponding semantic code regions.
These signals describe where readers concentrate their attention and the neural activity associated with reading individual regions or an entire program.

\paragraph{Model responses.} We process the same Java programs with 
each MoE model and record five responses at each candidate block: expert-choice concentration, expert-choice change, hidden-state shift, MoE write strength, and residual update strength.
Their exact definitions and aggregation are given in Section~\ref{sec:metrics}.
These quantities describe routing and representation changes at code locations.
We compare them with human measurements to examine correspondence, without assuming that response magnitude directly identifies an optimal adaptation location.

\paragraph{Alignment and preprocessing.} EEG is resampled to 100\,Hz, filtered at 0.5--40\,Hz, channel-screened, and common-average referenced.
4--8\,Hz theta is measured using median Hilbert log power across six nonfrontal regions, averaged over 0--1\,s after backward-fixation onset relative to a $-0.8$ to $-0.2$\,s baseline.
Local theta and dwell/regression/revisit summaries are standardized within participant and aggregated at mapped tokens.
Welch theta log-power changes relative to the $-5$ to $-1$\,s pre-code baseline are averaged across nine regions, standardized within participant and time bin, and averaged across participants and 12 normalized reading bins.
Fixations are mapped to visible code spans using image geometry and each model's tokenizer offsets.

Together, these aligned human and model measurements establish where human
processing demand corresponds with model computation on the same code.
However, downstream code-generation tasks do not come with EEG or eye-tracking
data. We therefore next summarize the human observations into transferable
guidance that can be applied to new training examples.

\subsection{Stage 2: Constructing Transferable Cognitive Priors} The human corpus contains Java programs, whereas downstream training uses Python solutions.
CogAdapt therefore transfers two summaries of the human observations: a program-level processing-demand prior and a token-level attention prior based on syntactic role.
These priors summarize patterns learned from the source corpus; they are not direct EEG or attention measurements of the downstream Python tasks.

\subsubsection{Program-Level Demand Prior}
\label{sec:program_difficulty}

Human EEG recordings provide a starting point for estimating the processing demand of reading code.
Prior work has used regression models to relate static code metrics to EEG-measured cognitive load~\cite{code_metrics_eeg2026}.
We follow this general strategy to transfer evidence from human code reading to downstream tasks.

For each of the Java programs, we summarize participant-normalized
theta responses across readers and the 12 reading-time bins.
We describe its structure using eight properties: character count,
nonblank-line count, lexical-token count, maximum nesting depth,
and counts of control-flow constructs, function calls, operators,
and statements~\cite{code_metrics_eeg2026}. A ridge regressor with standardized inputs and
regularization strength 10 learns how these properties relate to the program-level theta response.
For downstream task $i$, we extract corresponding properties
$\phi_i$ from its Python reference solution and compute
\begin{equation}
  d_i^{\mathrm{EEG}} = f_{\mathrm{EEG}}(\phi_i).
\end{equation}
Python tokenization and syntax trees provide approximate counterparts
to the Java features. 
Higher scores indicate resemblance to programs that elicited stronger
human theta responses. CogAdapt uses the standardized score to
weight training tasks in both models. The score is a proxy for reading demand; computing it
requires no EEG recordings for downstream tasks.

\subsubsection{Token-Level Attention Prior}
\label{sec:token_attention_prior}
Following prior work that uses
programmer visual attention to guide LLM training~\cite{zhang2026eyemulator,zhang2026scancoder}, we use attention behavior to
construct a transferable token-level attention prior.
For each code element observed in the human study, we summarize multiple
reading behaviors, including how long participants inspect the element and
whether they return to it during comprehension. We combine these signals into
an attention score $a_t$:

\begin{equation}
    a_t = f_{\mathrm{attention}}
    \left(
    \mathrm{dwell}_t,\,
    \mathrm{regression}_t,\,
    \mathrm{revisit}_t
    \right),
\end{equation}

where a larger $a_t$ indicates that the corresponding code element received
greater human attention.

Because the human data are collected on Java while our downstream tasks are written in Python, we do not transfer scores through exact token matching.
Instead, we transfer attention through syntactic correspondence.
We represent code elements using their syntactic roles and AST context and associate structurally similar elements across the two languages.
For a downstream token, its 
human-attention prior 
is therefore determined by the attention observed for syntactically similar elements in the human-study programs.
This design follows the idea established by prior approaches: human reading behavior is converted into reusable supervision for code-model training~\cite{zhang2024eyetrans,zhang2026eyemulator}.

\subsection{Stage 3: Selecting Blocks for Each Training Task}
\label{sec:block-selection}

Different coding tasks may benefit from adaptation in different parts of the model.
To identify these locations, we first pass each training prompt and its reference
solution through the pretrained MoE model. For each transformer block, we
measure how the MoE router changes its selection of routed experts, how much the
block changes its internal representation, and how strongly the selected experts
contribute to the block output. We summarize these responses at sampled solution
positions and normalize them within each task so that the different signals can be
compared on a common scale.
The previous analysis tells us which depth ranges tend to show stronger human–model correspondence. We summarize this pattern into a model-specific depth prior and use it as one source of evidence when ranking blocks for a new task.
The resulting ranking identifies the blocks favored by the selection policy for that particular task.
We also decide how much adaptation capacity to allocate.
Higher-demand tasks receive slightly more blocks, while lower-demand tasks receive fewer.
We compute it once before fine-tuning and reuse it whenever the task appears during training.

\subsection{Stage 4: cognition-informed Sparse Fine-Tuning}
\label{sec:sparse_finetuning}

After selecting the adaptation blocks for each task, CogAdapt fine-tunes the model using the human-derived priors from Stage 2.
These priors affect training at two complementary levels.
The program-level EEG prior determines how strongly an entire training example contributes to learning, while the attention prior determines which parts of its reference solution receive greater emphasis.
The task-specific mask from Stage 3 then determines where these gradients are allowed to update the model.

\paragraph{Emphasizing cognitively demanding tasks.} The program-level processing-demand score provides a transferable estimate of how demanding a training example may be for a human reader.
CogAdapt uses this signal to assign larger training weights to examples associated with greater predicted processing demand.

\paragraph{Emphasizing human-attended code.}
Within each reference solution, CogAdapt uses the attention prior from
Section~\ref{sec:token_attention_prior} to emphasize code elements that tend
to attract more human attention during comprehension.  We also
use syntactic context to preserve this emphasis when transferring the prior to
the downstream code-generation data.

Combining the task-level and token-level weights, the supervised objective for
task $i$ is

\begin{equation}
\mathcal{L}_i =
w_i^{\mathrm{task}}
\frac{1}{T_i}
\sum_{t=1}^{T_i}
w_{it}^{\mathrm{token}}
\left[
-\log P_{\theta}
\left(
y_{it}\mid x_i,y_{i,<t}
\right)
\right],
\label{eq:cognitive_objective}
\end{equation}

where $w_i^{\mathrm{task}}$ controls the importance of the overall training
example and $w_{it}^{\mathrm{token}}$ controls the importance of individual
code tokens. The underlying objective remains standard next-token prediction;
human data changes only how the supervision is distributed across examples
and code locations.

\paragraph{Restricting learning to selected blocks.}
CogAdapt applies the task-specific block mask obtained in Stage 3.
Although the full model participates in the forward pass, only the adaptation
parameters within the selected blocks are allowed to receive gradients for the
current example. The pretrained MoE experts remain frozen, while lightweight
attention adapters and router parameters in the selected blocks are updated.
Human signals determine
\emph{what training information should receive greater emphasis}, while the
task-dependent block mask determines \emph{where the resulting learning should
occur}. As a result, each example can focus both its supervision and its
adaptation capacity on the parts of the training process that are most relevant
to that task.

\subsection{Stage 5: Standard Inference without Human Signals}
\label{sec:inference}

After fine-tuning, CogAdapt is used in the same way as a regular code LLM.
Given a new programming prompt, the trained model directly generates the
solution using standard inference. No EEG, eye-tracking data,
reference solution, or additional cognitive information is required for the
downstream task.
\emph{The human measurements are used only during the earlier stages of CogAdapt to
construct transferable priors and guide how the model is adapted during
training.} Once training is complete, this information is already reflected in
the learned model parameters. As a result, CogAdapt does not require new human
data when applied to unseen code-generation problems, making the approach
compatible with standard deployment settings.

\section{Evaluation}
\label{sec:evaluation}

Our evaluation follows the progression from human--model correspondence to its practical use in model adaptation.
RQ1 first examines whether developers and MoE models respond more strongly to the same code regions.
RQ2 then examines where this correspondence appears across model depth and human reading stages.
These two analyses provide the empirical motivation for task-dependent block selection.
Finally, RQ3 evaluates whether using this information for cognition-informed sparse adaptation can improve code-generation accuracy while reducing the training resources required for adaptation.

We study Qwen3-Coder-30B-A3B-Instruct and GLM-4.7-Flash. Qwen contains
48 transformer blocks and routes each token to eight of 128 experts.
GLM contains 47 transformer blocks; each routed MoE block selects four
of 64 routed experts and additionally contains a shared expert. We
exclude GLM's shared expert when computing router-distribution statistics.

\subsection{Datasets and Benchmarks}
\label{sec:benchmarks}

\subsubsection{Human Code-Comprehension Corpus.} We use the NoviceVsExpert corpus~\cite{efficacy}, a controlled code-comprehension study with 37 participants.
Participants read 32 
Java programs of varying difficulty and answered questions about their behavior while their eye movements and EEG signals were recorded.
This provides complementary views of human code comprehension: reading time captures the overall effort spent on a program, eye tracking shows which parts of the code receive attention, and EEG provides a measure of neural activity during comprehension.
At finer granularity, we align eye fixations with the code being viewed and summarize how long readers look at each location, how often they revisit it, and when they look back to an earlier part of the program.
We normalize these attention measures within each participant before combining them across readers.
We further group nearby tokens into 81 semantic code regions, allowing human and model responses to be compared at the same meaningful code locations.

\subsubsection{Code-Generation Benchmarks}
\label{sec:generation_benchmarks}

We evaluate on two benchmarks: LiveCodeBench
(LCB)~\cite{livecodebench} and BigCodeBench (BCB)~\cite{bigcodebench}.
LCB supplies contest-style programming problems, while BCB tests
instruction following and the composition of library functions.
For LCB, the final training corpus contains 534 execution-validated
Python reference solutions prepared from versions 1--4; evaluation uses
175 tasks from the saved version-6 manifest.  For BCB, the saved
manifest partitions 1,140 tasks into 760 training, 160 development, and 220 test tasks.


\subsection{Metrics}
\label{sec:metrics}
\subsubsection{Model Computation Signals}
\paragraph{LLM Signal Extraction and notation.}
For token $n$ at block $l$, let $h^{\mathrm{in}}_n$ and
$h^{\mathrm{out}}_n$ be the block input and output, $a_n$ the attention
write, $m_n$ the complete MoE/MLP write, and
$u_n=h^{\mathrm{out}}_n-h^{\mathrm{in}}_n$ the total residual update.
We suppress the block index when defining individual metrics; $n-1$
always refers to the preceding token at the same block. Let $p_n$ denote
Qwen's normalized distribution over all 128 experts, $S_n$ its actual
top-eight dispatch set, and $w_n$ its captured dispatch-weight vector,
with zero entries for unselected experts. Full-distribution statistics
and actual dispatch statistics capture different aspects of routing.
GLM instead uses 64 routed experts and a shared expert; the shared
expert contributes to $m_n$ but is excluded from routing statistics.

\paragraph{Expert-choice confidence.}
Expert-choice confidence measures how strongly the router prefers a small set of experts for a
token. Intuitively, if the routing probability is spread nearly uniformly across experts, the router is
uncertain; if most of the probability is concentrated on a few experts, the router is more confident.
Let $p_e$ denote the normalized routing probability assigned to expert $e$, and let $E$ be the
number of routed experts. We use normalized routing entropy to illustrate this idea:
\begin{equation}
C_{\mathrm{conf}}
=
1 -
\frac{
-\sum_{e=1}^{E} p_e \log p_e
}{
\log E
}.
\label{eq:routing_confidence}
\end{equation}

$C_{\mathrm{conf}}$ is close to $0$ when the router distributes probability broadly across experts,
and approaches $1$ when its routing decision is concentrated on a small number of experts.
Thus, a larger value indicates a more decisive expert choice.

\paragraph{Expert-choice change.}
Expert-choice change measures how much the router changes its selected experts between two
neighboring code tokens. Intuitively, if two consecutive tokens use nearly the same experts,
the routing decision is stable; if they use substantially different experts, the model is reallocating
its computation.
Let $S_t$ and $S_{t-1}$ denote the sets of experts selected for the current and preceding token,
respectively. We illustrate this change using Jaccard distance:
\begin{equation}
C_{\mathrm{change}}
=
1 -
\frac{|S_t \cap S_{t-1}|}
{|S_t \cup S_{t-1}|}.
\label{eq:expert_change}
\end{equation}

$C_{\mathrm{change}}$ is close to $0$ when consecutive tokens use mostly the same experts,
and becomes larger as more experts are replaced. Thus, a larger value indicates a stronger
change in how the model allocates computation across experts.

\paragraph{Hidden-state shift.}
Hidden-state shift measures how much the model's internal representation changes between two
neighboring code tokens. Intuitively, if consecutive tokens are represented similarly, the shift is
small; if the next token substantially changes the representation, the shift is large.
Let $h_t$ and $h_{t-1}$ denote the hidden representations of the current and preceding token.
A simple measure of this change is the normalized Euclidean distance:
\begin{equation}
C_{\mathrm{hidden}}
=
\frac{\|h_t-h_{t-1}\|_2}
{\frac{1}{2}(\|h_t\|_2+\|h_{t-1}\|_2)}.
\label{eq:hidden_shift}
\end{equation}

A larger value therefore indicates that moving to the current token produces a stronger changein the model's internal representation.

\paragraph{MoE write.}
MoE write measures the strength of the contribution produced by the model's expert network at
the current token.
Let $m_t$ denote the output produced by the MoE/MLP component and $h_t^{\mathrm{in}}$ the
representation entering the block. A simple normalized measure is
\begin{equation}
C_{\mathrm{write}}
=
\frac{\|m_t\|_2}
{\|h_t^{\mathrm{in}}\|_2}.
\label{eq:moe_write}
\end{equation}

A larger value indicates that the experts make a stronger contribution relative to the
representation already present at that block.

\paragraph{Residual integration.}
Residual integration characterizes how strongly the computation inside a transformer block is
incorporated into the representation passed to the next block. Intuitively, it asks whether the
block makes only a small modification to the incoming representation or produces a substantial,
well-aligned update.

Let
\begin{equation}
u_t = h_t^{\mathrm{out}} - h_t^{\mathrm{in}}
\end{equation}
denote the total update produced by a block.
A simple measure of update strength is
\begin{equation}
C_{\mathrm{update}}
=
\frac{\|u_t\|_2}
{\|h_t^{\mathrm{in}}\|_2}.
\label{eq:residual_update}
\end{equation}

Larger values indicate that the block changes the incoming representation more strongly.

\subsubsection{Human--Model Correspondence}
\label{sec:human_model_correspondence}

\paragraph{Rank correlation.}
To measure whether humans and models respond strongly to the same code,
we use Spearman's rank correlation ($\rho$). Spearman correlation measures
whether two quantities follow a similar ordering without requiring them to
have the same scale or a linear relationship~\cite{schober2018correlation}.
This is useful in our setting because human measurements, such as attention and
EEG, and model-computation signals have very different units.

Formally, we compute

\begin{equation}
\rho(X,Y)=\operatorname{corr}\!\left(
\operatorname{rank}(X),\operatorname{rank}(Y)\right),
\label{eq:spearman}
\end{equation}

where $X$ is a human response and $Y$ is a model response measured over the
same observations. A positive $\rho$ means that code locations or programs
receiving stronger human responses also tend to produce stronger model
responses; a negative value indicates the opposite ordering, while values
closer to zero indicate less consistent correspondence. Depending on the
analysis, each observation represents a token, a semantic code region, or a
complete program. This type of human--model comparison follows prior work
showing that representations of source code can exhibit correspondence
between human brain activity and learned code representations
~\cite{srikant}.

\paragraph{Depth--time correspondence.}
We further ask whether human--model correspondence is distributed across
the model and whether it changes as humans progress through a program.
Prior studies of language processing have shown that brain--model
correspondence can vary across both model layers and stages of human
processing~\cite{caucheteux2022,schrimpf}. Following this
idea, we compare each model-computation signal at every transformer block
with human EEG theta measured over 12 normalized stages of program reading.
For model signal $s$, block $b$, and reading stage $t$, we summarize this
relationship as

\begin{equation}
\rho^{(s)}_{b,t}
=
\rho\!\left(
M^{(s)}_{b},\,
\Theta_t
\mid \mathrm{Length}
\right),
\label{eq:depth_time}
\end{equation}

where $M^{(s)}_{b}$ is the model response across programs and $\Theta_t$ is
the corresponding human theta response. We control for program length so
that the comparison reflects more than simply larger programs producing
larger responses.
The resulting depth--time map provides an intuitive view of the relationship:
each cell asks whether, at a particular model block and stage of human
reading, programs that elicit stronger human responses also elicit stronger
model computation.

\subsubsection{Execution Accuracy and Statistical Comparison}
\label{sec:execution_metrics}

We assess code-generation quality on the held-out tasks described in
Section~\ref{sec:generation_benchmarks} using execution-based pass@1. 
Pass@1 is a percentage of solved tasks; $\Delta_{\mathrm{pass}}$ is an
absolute percentage-point difference. Each condition uses its first
complete generation pass. For Random-K6, the reported standard deviation is across three
fixed masks. 

\subsubsection{Adaptation Sparsity and Training Efficiency.}
We evaluate how selectively \textsc{CogAdapt} allocates adaptation capacity and whether this selectivity translates into lower training cost.
\textbf{Mean selected blocks} reports the average number of transformer blocks selected for adaptation for each unique training task. 
\textbf{Gradient-eligible parameters} measures the average number of attention-adapter and router parameters that are allowed to receive gradients for one task. 
\textbf{Adaptation reduction} reports the percentage reduction in these gradient-eligible parameters relative to the all-block Regular FT baseline. 

To measure practical training cost, we additionally report \textbf{training wall time} and \textbf{GPU energy consumption}. Wall time records the elapsed training time, while GPU energy is estimated from sampled GPU power over the recorded measurement interval. We report their reductions relative to the matched Regular FT condition, where a positive reduction indicates lower measured training cost.

\subsection{Baselines and Ablations}
\label{sec:baselines}

We compare CogAdapt with standard and sparse fine-tuning baselines, and
use a set of controlled ablations to examine where its gains come from.
Unless otherwise stated, all methods use the same training data,
optimization settings, LoRA configuration, and evaluation protocol.

\subsubsection{Baselines}

\paragraph{Regular fine-tuning (Regular FT).}
This baseline applies the same attention-LoRA and router-gate adaptation
to every candidate block. It uses the same benchmark-specific training
data, adapter rank, training exposure, and optimization settings as
CogAdapt, but does not perform selective block adaptation.

\paragraph{Representation-based block selection (CKA-K6).}
We further compare with a model-internal sparse selection strategy based
on centered kernel alignment (CKA), following prior work on using
representation similarity to identify transformer layers for
adaptation~\cite{ogawa2026layerwise}. For each block, we compute linear
CKA between its input and output representations, extracted at the final
non-padding token of the training prompts and reference solutions. For
centered representation matrices $X_c$ and $Y_c$,
\begin{equation}
\operatorname{CKA}(X,Y)=
\frac{\|X_c^{\mathsf T}Y_c\|_F^2}
{\|X_c^{\mathsf T}X_c\|_F
 \|Y_c^{\mathsf T}Y_c\|_F},
\label{eq:cka}
\end{equation}
where $\|\cdot\|_F$ denotes the Frobenius norm.
A lower CKA score indicates a larger representational change between a
block's input and output. We therefore select the six blocks with the
lowest CKA scores and keep this set fixed during adaptation. Unlike
CogAdapt, this baseline determines where to fine-tune entirely from the
model's internal representations, without using human cognitive signals.

\subsubsection{Ablations}

We design five ablations to isolate the main components of CogAdapt.
They differ primarily in \textbf{which human signals are used} and
\textbf{how adaptation blocks are selected}. Unless otherwise stated,
all variants use the same training data, optimization settings, and
evaluation protocol as CogAdapt.

\paragraph{All-Block.}
This variant retains CogAdapt's EEG-derived task weighting and
attention-derived token weighting, but \textbf{updates all candidate
blocks instead of selectively chosen blocks}.

\paragraph{Rand-K6.}
This variant retains CogAdapt's human-derived task and token weighting,
but \textbf{replaces cognitive-informed block selection with six
randomly selected blocks}. The same six blocks are fixed across all
tasks and throughout training and inference.

\paragraph{Model-Only.}
This variant \textbf{removes all human-derived guidance}. For each task,
it selects blocks only from the model's own behavior. The
highest-scoring blocks are then fine-tuned, so different tasks can adapt
different blocks. 

\paragraph{No-EEG.}
This variant \textbf{removes EEG-derived task weighting and EEG-based
block-selection guidance}, while retaining attention based
token weighting.

\paragraph{No-Gaze.}
This variant \textbf{removes attention-based token weighting}, so all
supervised completion tokens receive equal loss weight. EEG-derived task
weighting and block selection remain unchanged.

\subsection{Training and Hardware Settings}
\label{sec:training_hardware}

\paragraph{Training configuration.}
Attention adapters use rank 16,
scale 32, and dropout 0.05. AdamW uses learning rates $7\times10^{-5}$
for attention adapters and $1.5\times10^{-5}$ for router gates,
$\beta_1=0.9$, $\beta_2=0.999$, $\epsilon=10^{-8}$, and weight decay
0.01. Learning rates follow a 3\% linear warmup and cosine decay;
the gradient norm is clipped at 1.0. The microbatch contains one example,
with eight microsteps accumulated per update. 
The recorded setup uses two NVIDIA RTX A6000 GPUs. The reproduction host exposes an Intel Xeon Gold 6330N CPU
at 2.20 GHz, 28 physical cores, 56 logical CPUs, and 256GiB of RAM.

\section{Results}
\label{sec:results}


\begin{figure*}[t]
    \centering
    \includegraphics[width=\textwidth]
    {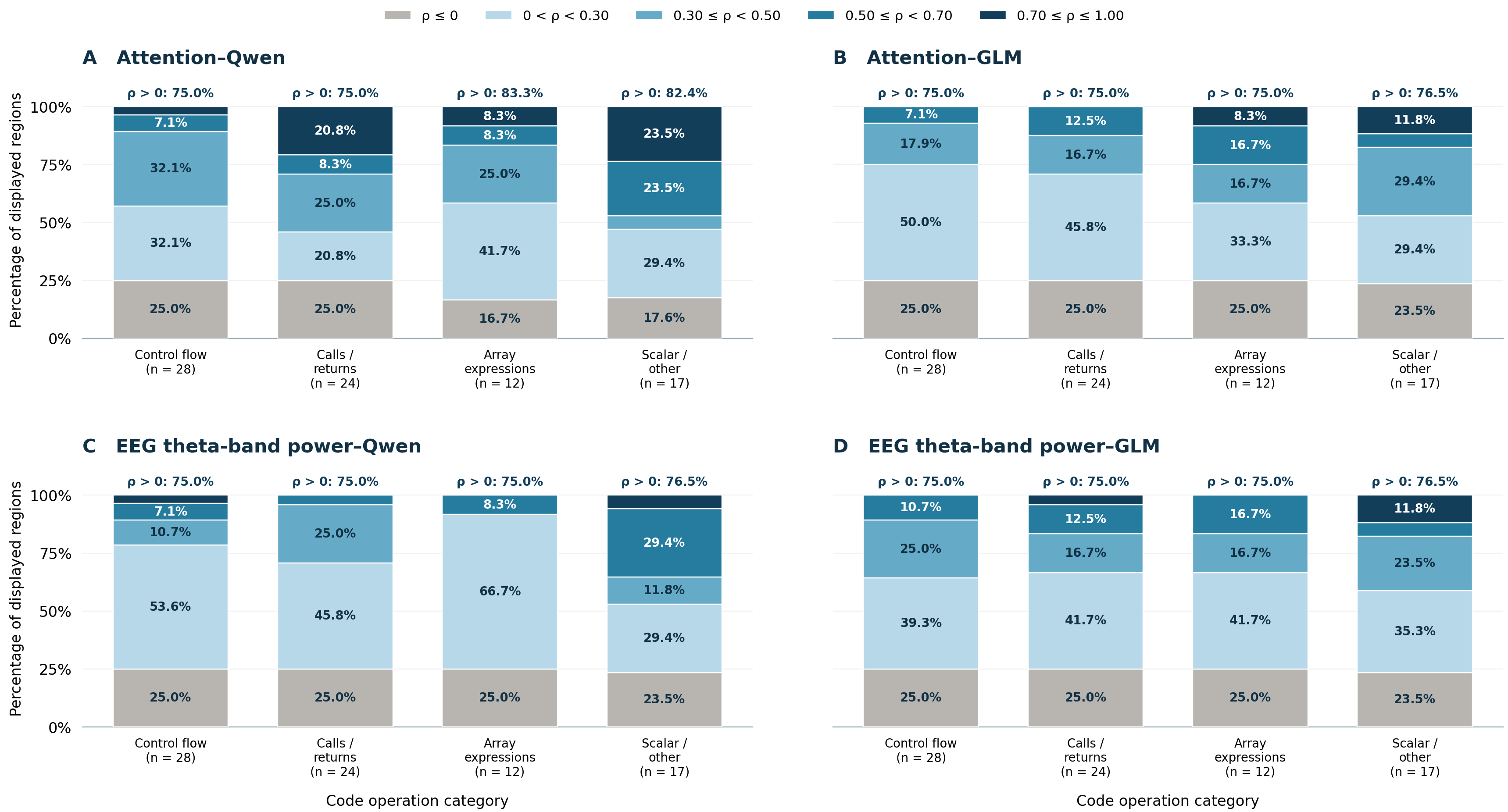}
\caption{\textbf{Human–model alignment across code categories.} The x-axis groups code regions into four semantic categories. The y-axis shows the percentage of regions in each correlation range. Gray indicates nonpositive correlations; increasingly dark blue indicates stronger positive correlations. The four panels compare human attention or EEG theta-band activity with Qwen or GLM. For each region, Spearman correlation is calculated across matched code tokens between the human signal and the mean of five standardized model metrics. Across all displayed regions, median correlations and approximate
95\% resampling intervals are:
Attention--Qwen, $0.240$ [$0.084$, $0.341$];
Attention--GLM, $0.200$ [$0.071$, $0.288$];
EEG theta--Qwen, $0.183$ [$0.000$, $0.217$]; and
EEG theta--GLM, $0.200$ [$0.011$, $0.214$].
Intervals use 5,000 resampling draws accounting for repeated
observations from the same participants and programs.}
\label{fig:rq1_region_heatmap}
\end{figure*}

\subsection{RQ1: Do Humans and MoE Models Respond More Strongly to the Same Code Regions?}
\label{sec:rq1}

\paragraph{Human reading responses and MoE computation are positively correlated across code categories.}
Figure~\ref{fig:rq1_region_heatmap} provides the main view of this
relationship. We group semantic code regions into four common
categories---control flow, calls/returns, array expressions, and
scalar/other operations---and examine the distribution of Spearman
correlations within each category. The four panels compare human
attention and EEG theta-band activity with the model responses of Qwen
and GLM. Across all human-signal/model combinations and all four code
categories, \textbf{75.0\%--83.3\% of the displayed regions show
positive correlations} ($\rho>0$). In other words, code regions that
receive stronger human responses also tend to induce stronger model
computation.

The correspondence is also \textbf{not uniform across code regions}.
Some regions show relatively weak positive relationships, whereas
others exhibit stronger correspondence. For example, Attention--Qwen
correspondence is positive for \textbf{83.3\% of array-expression
regions} and \textbf{82.4\% of scalar/other regions}, while the other
category/model combinations remain positive for roughly three quarters
or more of the regions. The strength of the relationship also differs
between Qwen and GLM and between attention and EEG, suggesting that
\textbf{human-salient code is reflected through different internal
responses across model architectures}.

\paragraph{Answer to RQ1.}
\emph{\textbf{Human reading responses and MoE computation show positive correspondence at the same semantic code
locations.}} This pattern appears across both attention and EEG and
across both model families, although its strength varies by code region
and architecture. Importantly, this result provides a practical
motivation for CogAdapt's training design: \textbf{human cognitive
signals can identify code regions that are also computationally salient
to the model}, supporting their use as guidance for emphasizing
training examples and code tokens during adaptation.

\begin{figure*}[t]
    \centering
    \includegraphics[width=\textwidth]
    {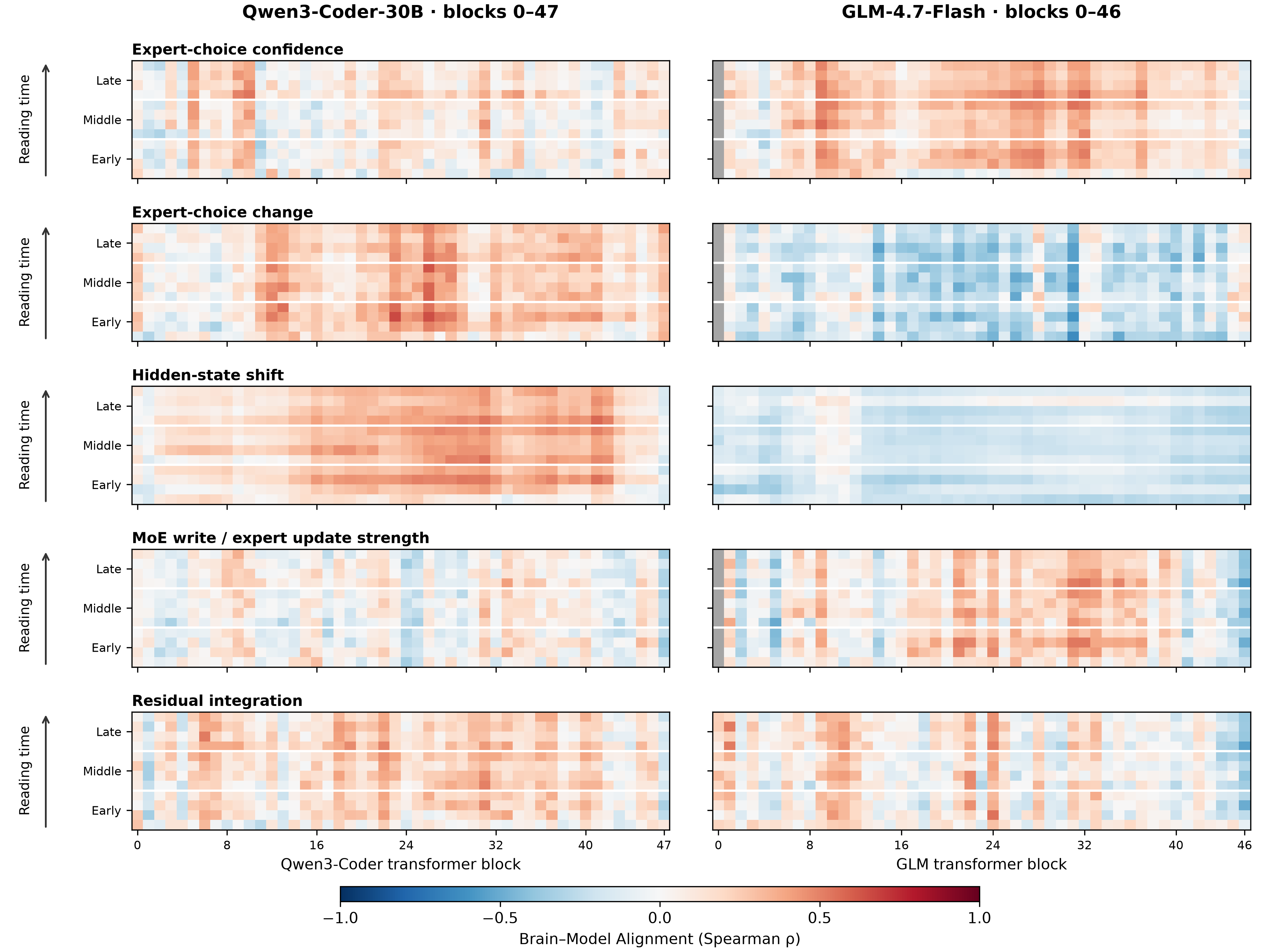}
    \caption{\textbf{Human brain–model alignment during code reading.} The x-axis shows successive transformer blocks (model layers), from earlier to deeper processing. The y-axis shows human reading time, progressing upward from Early through Middle to Late. Colors show how each model measure relates to human EEG theta activity across the same 32 programs. Red indicates a positive association, blue a negative association.}
    \label{fig:theta_depth_time}
\end{figure*}
\begin{figure*}[t]
    \centering
    \includegraphics[width=\textwidth]
    {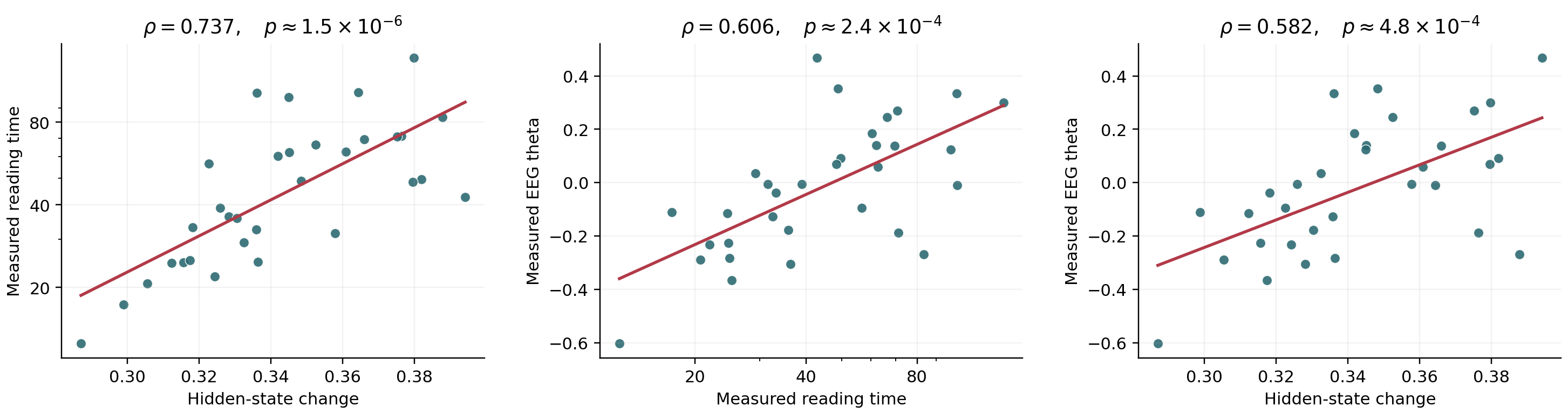}
\caption{An example of program-level human--model correspondence at Qwen block~41. Each point represents one of the 32 programs. The left panel shows that programs producing larger hidden-state changes in the model tend to require longer human reading time. The middle panel shows that longer reading time is associated with stronger EEG theta activity, while the right panel shows a direct positive relationship between model hidden-state change and EEG theta activity.}
\label{fig:router_rt_theta}
\end{figure*}

\subsection{RQ2: Where in the MoE Model Does Computation Correspond to Human Processing Demand?}
\label{sec:rq2}

\paragraph{Human--model correspondence forms structured patterns across model depth.}
Figure~\ref{fig:theta_depth_time} provides the main result for RQ2. For
each transformer block, we compare five model-computation signals with
human EEG theta activity across the 32 programs, while also considering
when the EEG response occurs during the reading process. Each cell
reports a Spearman correlation: warmer regions indicate that programs
producing stronger model responses also tend to elicit stronger human
theta responses, whereas cooler regions indicate the opposite
relationship.

The heatmaps reveal positive correspondence, but importantly, this
correspondence is \textbf{structured rather than uniform}. For Qwen,
\textbf{hidden-state shift shows broad positive associations across
middle and later blocks} and remains visible across multiple stages of
human reading. Expert-choice change also develops extended positive
regions through the middle and deeper portions of the model, while
residual integration is frequently positive across a broad span of
blocks. Expert-choice confidence and MoE write show more localized
positive regions. Together, these patterns indicate that
\textbf{human--model correspondence is stronger at some model depths
than at others}, rather than being distributed uniformly across the
network.

GLM exhibits a different, but similarly organized, pattern.
\textbf{Expert-choice confidence and MoE write show the clearest
positive correspondence with EEG theta}, particularly across middle
and deeper blocks. In contrast, hidden-state shift and expert-choice
change are largely negative, while residual integration contains a
mixture of positive and negative regions. Thus, the two architectures
do not express human--model correspondence through the same internal
quantity: \textbf{Qwen shows clearer correspondence in representational
change and integration, whereas GLM shows clearer correspondence in
routing confidence and expert contribution}. The important commonality
is that both models contain identifiable ranges of blocks where human
processing demand corresponds systematically with model computation.

\paragraph{The correspondence is distributed rather than tied to one special block.}
Figure~\ref{fig:router_rt_theta} provides a concrete example of the
broader pattern in Figure~\ref{fig:theta_depth_time}. For Qwen block 41,
programs with larger hidden-state changes also tend to require longer
human reading time. Human reading time is itself
positively associated with EEG theta activity,
while hidden-state change is directly correlated with EEG theta activity. This example illustrates that programs requiring
greater human processing can also produce greater changes in the model's internal computation.

We do not interpret block 41 as a uniquely ``brain-aligned'' block.
Rather, Figure~\ref{fig:theta_depth_time} shows that related positive
associations occur across multiple neighboring depths, and that their
locations and computational forms differ between model architectures.
The key finding is therefore that \textbf{the correspondence is
distributed, depth-dependent, and architecture-dependent}. If
human-related computation were concentrated in one universal block, a
fixed adaptation location could be sufficient. Instead, the relevant
blocks vary across the model and across architectures. This directly
motivates CogAdapt's task-dependent design: \textbf{rather than updating
all blocks or using one fixed sparse subset, CogAdapt selects different
blocks for different coding tasks}, while human-derived processing
signals guide the amount and emphasis of adaptation.

\paragraph{Answer to RQ2.}
\emph{\textbf{Human processing demand exhibits systematic correspondence
with MoE computation across model depth.}} Figure~\ref{fig:theta_depth_time}
shows broad and structured positive regions in both model families,
although the model signals carrying this correspondence differ between
Qwen and GLM. The relationship is therefore \textbf{distributed,
depth-dependent, and architecture-dependent}, rather than confined to a
single universally aligned block.  This provides the empirical basis for
CogAdapt: \textbf{human signals reveal where meaningful model
computation occurs, while the lack of one universally optimal block
motivates dynamically selecting adaptation blocks for each coding task.}

\begin{table*}[t]
\centering
\caption{
Pass@1 (\%) of baselines, CogAdapt ablations, and the full method.
Rand-K6 reports mean $\pm$ standard deviation over three fixed random
six-block masks. Bold indicates the best result in each 
setting. 
}
\label{tab:accuracy_ablation}
\small
\setlength{\tabcolsep}{4.5pt}
\resizebox{\textwidth}{!}{
\begin{tabular}{llcc|ccccc|c}
\toprule
&
&
\multicolumn{2}{c|}{\textbf{Baselines}}
&
\multicolumn{5}{c|}{\textbf{Ablations}}
&
\textbf{Full} \\
\cmidrule(lr){3-4}
\cmidrule(lr){5-9}
\cmidrule(lr){10-10}

\textbf{Model}
&
\textbf{Benchmark}
&
\textbf{Regular FT}
&
\textbf{CKA-K6}
&
\textbf{All-Block}
&
\textbf{Rand-K6}
&
\textbf{Model-Only}
&
\textbf{No-EEG}
&
\textbf{No-Gaze}
&
\textbf{CogAdapt}
\\
\midrule

Qwen
& LCB
& 18.86
& 24.57
& 20.00
& $22.86 \pm 0.57$
& 24.00
& 26.29
& 28.57
& \textbf{29.71}
\\

Qwen
& BCB
& 36.36
& 32.73
& 34.55
& $34.39 \pm 1.14$
& 36.36
& 35.91
& \textbf{37.73}
& \textbf{37.73}
\\

\midrule

GLM
& LCB
& 17.71
& 21.14
& 20.00
& $20.00 \pm 1.51$
& 22.29
& 21.14
& 21.71
& \textbf{24.00}
\\

GLM
& BCB
& 28.64
& 28.64
& 30.91
& $30.76 \pm 0.95$
& 28.18
& 29.55
& 30.00
& \textbf{31.82}
\\

\bottomrule
\end{tabular}
}
\end{table*}

\begin{table*}[t]
\centering
\caption{
Training-side efficiency of task-dependent sparse adaptation.
}
\label{tab:cost}
\small
\setlength{\tabcolsep}{4.5pt}
\renewcommand{\arraystretch}{1.15}
\resizebox{\textwidth}{!}{
\begin{tabular}{llccccccccc}
\toprule

\textbf{Model} &
\textbf{Bench.} &
\shortstack{\textbf{Mean}\\\textbf{Blocks}} &
\shortstack{\textbf{Eligible}\\\textbf{Params (M)}} &
\shortstack{\textbf{All-block}\\\textbf{Params (M)}} &
\shortstack{\textbf{Adapt.}\\\textbf{Reduction}} &
\shortstack{\textbf{Train Time (s)}\\\textbf{Ours / FT}} &
\shortstack{\textbf{Time}\\\textbf{Reduction}} &
\shortstack{\textbf{GPU Energy (Wh)}\\\textbf{Ours / FT}} &
\shortstack{\textbf{Energy}\\\textbf{Reduction}} \\

\midrule

Qwen &
LCB &
6.25 / 48 &
3.382 &
25.952 &
\textbf{86.97\%} &
1883.78 / 1944.82 &
\textbf{3.14\%} &
188.10 / 192.81 &
\textbf{2.44\%} \\

Qwen &
BCB &
6.14 / 48 &
3.319 &
25.952 &
\textbf{87.21\%} &
2201.15 / 2304.37 &
\textbf{4.48\%} &
190.02 / 197.04 &
\textbf{3.56\%} \\

\midrule

GLM &
LCB &
6.09 / 47 &
3.486 &
27.061 &
\textbf{87.12\%} &
1608.03 / 1649.79 &
\textbf{2.53\%} &
168.99 / 172.66 &
\textbf{2.12\%} \\

GLM &
BCB &
6.55 / 47 &
3.733 &
27.061 &
\textbf{86.21\%} &
1888.42 / 2006.22 &
\textbf{5.87\%} &
174.00 / 182.28 &
\textbf{4.55\%} \\

\bottomrule
\end{tabular}
}
\end{table*}

\subsection{RQ3: Does CogAdapt Improve Code Generation with More Selective Adaptation?}
\label{sec:rq3}

\paragraph{CogAdapt improves accuracy over both dense and fixed sparse adaptation.}
Table~\ref{tab:accuracy_ablation} shows that CogAdapt achieves the best
or tied-best pass@1 across the four model--benchmark settings.
The improvement is especially clear on LiveCodeBench. For Qwen,
pass@1 increases from 18.86\% with Regular FT to \textbf{29.71\%}
with CogAdapt, a gain of \textbf{10.86 percentage points}, while GLM
improves from 17.71\% to \textbf{24.00\%}, a gain of
\textbf{6.29 points}. On BigCodeBench, CogAdapt reaches
\textbf{37.73\%} for Qwen and \textbf{31.82\%} for GLM, again
outperforming Regular FT. CogAdapt also exceeds the fixed CKA-K6
baseline in all four settings, suggesting that task-dependent adaptation
is more useful than committing to the same representation-based block
set for every task. 

\paragraph{Ablations show that both block selection and human guidance matter.}
Simply applying human-weighted training to every block does not produce
the same benefit: All-Block performs below CogAdapt in all four settings.
Replacing informed block selection with a fixed random set also reduces
accuracy, with Rand-K6 trailing CogAdapt across both models and
benchmarks. Model-Only, which chooses blocks from the model's own
responses but removes human guidance, also performs consistently below
the full method. For example, its pass@1 drops from \textbf{29.71\% to
24.00\%} on Qwen-LCB and from \textbf{24.00\% to 22.29\%} on GLM-LCB.
These results indicate that CogAdapt benefits not only from adapting
fewer blocks, but also from using human signals to guide where and how
adaptation occurs.
The individual cognitive signals provide complementary contributions.
Removing EEG reduces Qwen-LCB from 29.71\% to \textbf{26.29\%},
Qwen-BCB from 37.73\% to \textbf{35.91\%}, and GLM-LCB from
24.00\% to \textbf{21.14\%}. Removing gaze has a smaller effect on
Qwen-LCB, reducing pass@1 to \textbf{28.57\%}, while GLM-LCB drops
from 24.00\% to \textbf{21.71\%}. On Qwen-BCB, No-Gaze matches
CogAdapt at 37.73\%.

\paragraph{CogAdapt concentrates learning in only a small part of the model.}
Table~\ref{tab:cost} shows that CogAdapt selects only about
\textbf{6.1--6.6 blocks per task}, compared with all 47--48 blocks
under Regular FT. This reduces gradient-eligible adaptation parameters
from 25.95--27.06M to only \textbf{3.32--3.73M}, corresponding to an
\textbf{86.21--87.21\% reduction}. Despite executing the full backbone
during the forward pass, CogAdapt also reduces wall-clock training time
by \textbf{2.53--5.87\%} and measured GPU energy by
\textbf{2.12--4.55\%}. 

\paragraph{Answer to RQ3.}
\textit{CogAdapt achieves stronger or comparable code-generation
accuracy while adapting only about \textbf{six blocks per task}.
Its largest improvements over Regular FT are \textbf{10.86} and
\textbf{6.29 percentage points} on LiveCodeBench, while reducing
gradient-eligible adaptation parameters by approximately
\textbf{86--87\%}. The ablations further show that both informed block
selection and human cognitive guidance contribute to these gains.}

\section{Discussion, Limitations, and Future Directions}

Our results show that cognition-informed sparse adaptation can improve code generation while reducing the number of parameters updated during training. We next discuss why this selective adaptation may be effective, its current computational limitations, and directions
for extending sparsity beyond model adaptation.

\paragraph{Why selective adaptation can improve accuracy.} CogAdapt's gains suggest that adaptation quality depends on where updates occur, not only on how many parameters receive gradients.
Model responses identify blocks relevant to each task, while human-derived difficulty priors control the adaptation budget.
Restricting each example's updates to these blocks may reduce interference with representations useful for other tasks, while preserving the pretrained expert weights.
Human-derived task and token weights further concentrate supervision on examples and code regions associated with greater processing demand.
Together, these mechanisms may help a smaller per-example adaptation budget produce more useful updates.
CogAdapt's improvements over both all-block fine-tuning and fixed sparse baselines support the value of task-dependent selection.

\paragraph{From sparse adaptation toward sparse computation.} CogAdapt reduces the number of parameters that are eligible for gradients for each training example, but the full backbone still participates in the forward pass.
Consequently, the reduction in wall-clock time and energy is smaller than the reduction in adaptation parameters.
This distinction suggests an important next step: connecting cognitively informed adaptation with conditional model execution.
For example, future systems could extend sparsity beyond standard MoE
routing. While an MoE model already selects only a subset of experts within
each MoE block, the full sequence of transformer blocks still participates in
the forward pass. A future system could additionally use human-informed
signals to decide which blocks or adapters should participate for a given
task, or whether some block-level computation can be skipped entirely.
Combining expert routing with conditional block or adapter activation, early
exiting, or other forms of dynamic execution could therefore translate
CogAdapt's adaptation sparsity into larger training and inference savings.

\section{Conclusion}
\label{sec:conclusion}

We present CogAdapt, a cognition-informed framework that uses human attention and EEG signals to guide task-dependent sparse adaptation of MoE code models.
Our analysis shows that human processing and model computation correspond both at individual code regions and across model depth, although the specific patterns differ between architectures.
Rather than assuming a single universally aligned layer, CogAdapt transfers these cognitive signals as priors to guide what training examples and code regions receive more emphasis and where adaptation occurs within the model.

Across two MoE backbones and two code-generation benchmarks, CogAdapt
achieves the best pass@1 among the evaluated methods.
Compared with regular fine-tuning, it improves LiveCodeBench
pass@1 by 10.86 percentage points for Qwen and 6.29 points for GLM.
Our ablations further show that these gains are not explained by sparse
fine-tuning alone: using all blocks, fixed random blocks, or model-only
block selection generally performs worse than the full CogAdapt, while
removing EEG or attention guidance also reduces performance in most
evaluated settings. At the same time, CogAdapt reduces per-example
gradient-eligible adaptation parameters by 86.21--87.21\%, together
with reductions in training time and estimated GPU energy.
Overall, our findings show that human comprehension signals can provide
useful guidance not only for \emph{what} a model should learn from, but
also for \emph{where} learning should occur.

\bibliographystyle{ACM-Reference-Format}
\bibliography{related_work,introduction}

\end{document}